\documentclass[12pt,preprint]{aastex}
\usepackage{graphicx}
\usepackage{natbib}

\newcommand{\kms}{\hbox{km~s$^{-1}$}}

\newcommand{\as}{$^{\prime\prime}$}

\newcommand{\msun}{\ensuremath{M_{\odot}}}

\begin{document}

\title{Kinematically Identified Recoiling Supermassive Black Hole Candidates in SDSS QSOs with z $<$ 0.25}

\author{
D.-C. Kim\altaffilmark{1}, 
A. S. Evans\altaffilmark{1, 2},
S. Stierwalt\altaffilmark{1}, \&
G. C. Privon\altaffilmark{3, 4}
}
\altaffiltext{1}{National Radio Astronomy Observatory, 520 Edgemont Road,
Charlottesville, VA 22903: dkim@nrao.edu, aevena@nrao.edu, sstierwa@nrao.edu}

\altaffiltext{2}{Department of Astronomy, University of Virginia, 530
McCormick Road, Charlottesville, VA 22904}

\altaffiltext{3}{Departamento de Astronom{\' i}a, Universidad de Concepci{\' o}n, Concepci{\' o}n, Chile}
\altaffiltext{4}{Instituto de Astrof{\' i}sica, Facultad de F{\' i}sica, Pontificia Universidad Cat{\' o}lica de Chile, 306, Santiago 22, Chile: gprivon@astro.puc.cl}

\begin{abstract}
We have performed a spectral decomposition to search for recoiling supermassive black holes (rSMBH) in the SDSS QSOs with $z<0.25$.
Out of 1271 QSOs, we have identified 26 rSMBH candidates that are recoiling toward us. 
The projected recoil velocities range from $-76\ \kms$ to $-307\ \kms$ with a mean of $-149\pm58\ \kms$.
Most of the rSMBH candidates are hosted by gas-rich LIRGs/ULIRGs, but only 23\% of them shows signs of tidal features suggesting majority of them are advanced mergers.
We find that the black hole masses $M_{BH}$ of the rSMBH candidates are on average $\sim$5 times smaller than 
that of their stationary counterparts and cause a scatter in $M_{BH}-\sigma_*$ relation.
The Eddington ratios of all of the rSMBH candidates are larger than 0.1, with mean of 0.52$\pm$0.27, suggesting they are actively accreting mass.
Velocity shifts in high-excitation coronal lines suggest that the rSMBH candidates are recoiling with an average velocity of about $-265\ \kms$.
Electron density in the narrow line region of the H II rSMBH candidates is about 1/10 of that in AGN rSMBH candidates 
probably because AGN in the former was more spatially offset than that in the latter.
The estimated spatial offsets between the rSMBH candidate and center of host galaxy 
range from 0.21\as \ to 1.97\as \ and need to be confirmed spatially with high-resolution adaptive optics imaging observations.

\end{abstract}


\section{Introduction}

Galaxy interaction plays an important role in the evolution of galaxies;
it enhances strong starbust activities (Larson \& Tinsley 1978; Joseph et al. 1984; Sanders et al. 1988), 
induces nuclear and AGN feedback (Lehnert \& Heckman 1996; Heckman et al. 2000; Rupke et al. 2002), 
enriches the intergalactic medium with outflows (Nath \& Trentham 1997; Scannapieco et al. 2002),
and grows the mass of stellar bulges and supermassive black holes (SMBHs) (Kormendy \& Richstone 1995; Magorrian et al. 1998; Gebhardt et al. 2000; Ferrarese \& Merritt 2000; Tremaine et al. 2002; Marconi \& Hunt 2003; Hopkins et al. 2005).
When two SMBHs coalesce at the final stage of galaxy interaction,
{\it a merged SMBH} can recoil from the host galaxy due to anisotropic emission of gravitational waves.
Recent simulations of merging black holes predict that the merged SMBH can attain a kick velocity of
a few hundred to a few thousands km s$^{-1}$ depending on mass ratios, spin magnitudes,
and spin orientations of the merging SMBHs (Campanelli et al. 2007; Schnittman 2007; Baker et al. 2008; Lousto \& Zlochower 2011).
If the merging SMBHs are of equal mass, are highly-spinning, and their spins are aligned along the orbital plane (superkick configuration), 
recoil velocities as large as 4000 km s$^{-1}$ (Campanelli et al. 2007) to 5000 km s$^{-1}$ (Lousto \& Zlochower 2011) can be reached
and the recoiling SMBH (hereafter rSMBH) will eventually escape from the host galaxy (i.e., Merritt et al. 2004).

The rSMBH carries along with it the broad-line region (BLR) and leaves the starburst nucleus behind (Blecha et al. 2011; Sijacki et al. 2011).
Thus, they will have two observational characteristics: 
i) broad emission lines will have velocity offsets with respect to the systemic velocity (Merritt et al. 2006; kinematic offset), and 
ii) the center of the BLR will be spatially offset from stellar center of the host galaxy (spatial offset).
Bonning et al. (2007) conducted a search for rSMBHs by measuring kinematic offsets in the Sloan Digital Sky Survey (SDSS) DR3 QSO spectra,
but failed to detect any.
They placed upper limits on the incidence of rSMBHs in QSOs
of 0.2\% for kick velocities greater than 800 \kms.
Their result suggests that the majority of the kick velocities might be a few hundreds \kms\ or even less than 100 \kms\ 
(i.e. Dotti et al. 2010). In such a case, broad emission lines from rSMBHs will be well
blended with the narrow emission lines, and a careful deblending is required to decompose each emission line component.
A similar systematic search was carried out by Eracleous et al. (2012), but mainly aimed at detecting close binary SMBHs
whose broad H$\beta$ emission lines are displaced from the quasar rest frame by more than 1000 \kms. 
By analyzing Hubble Space Telescope archival images of 14 nearby core ellipticals,
Lena et al. (2014) find evidence for small (10 pc) displacements between the AGN and host galaxy center in six of them.
They found that the displacement axis in 4 out of these 6 galaxies is aligned with the radio source axis 
suggesting intrinsic asymmetries in radio jet power as a possible displacement mechanism.
Comerford \& Greene (2014) surveyed SDSS spectra 
to search for offset AGNs that are produced by galaxy interaction 
where one of the black holes is fueled as an AGN, 
but they excluded sources with velocity offsets produced by rSMBHs. 

One of the interesting rSMBH candidates so far is a CID-42 (also known as CXOC J100043.1+020637).
It has two compact optical sources separated by $\sim$2.5 kpc and has a large velocity offset ($\sim$1200 $\kms$) 
between the H$\beta$ narrow and broad emission line components (Civano et al. 2010).
Chandra observations find that the south-eastern source that has a point-like morphology typical of a bright
AGN is responsible for the whole X-ray emission in this system (Civano et al. 2012).
The north-western source has a more extended profile in the optical band with a scale length of $\sim$0.5 kpc.
Recent 3 GHz Karl G. Jansky Very Large Array (VLA) observations find that the entire observed 3 GHz radio emission can be
associated with the south-eastern component of this system (Novak et al. 2015).
Their finding preferred rSMBH picture but cannot rule out the presence of an obscured and radio-quiet SMBH in the north-western source.

Detection of rSMBHs will have a significant impact in astrophysics.
A scaling law between the SMBH mass and host galaxy bulge has been suggested by many correlations:
SMBH mass vs. bulge luminosity (Kormendy \& Richstone 1995; Marconi \& Hunt 2003),
bulge mass (Magorrian et al. 1998),
and stellar velocity dispersion (Gebhardt et al. 2000; Ferrarese \& Merritt 2000; Tremaine et al. 2002).
These correlations implicitly assume that the merged SMBH stays in the center of host galaxy.
If the recoil velocity is less than the escape velocity ($v_{\rm e}\approx$1500 km s$^{-1}$ in elliptical galaxies, i.e. Merritt et al. 2004), which is true in most cases (Dotti et al. 2010),
the rSMBHs will be oscillating about the center of their host galaxies (Blecha et al. 2011).
From numerical simulations, Blecha et al. (2011) predicted that the oscillating SMBHs may be up to five times less massive than their
stationary counterparts and could be a source of intrinsic scatter in the SMBH and stellar bulge mass scaling law.
The AGN feedback and outflow regulate starburst activity in merging galaxies, 
but if the central AGN is displaced by a recoil event, it enhances star formation rates in the center of host galaxy (Blecha et al. 2011).
The oscillating SMBHs can also induce interesting results in unified models of AGN.
An obscured type 2 QSO with its BLR hidden by the molecular torus will appear as unobscured type 1 QSO during the recoil
oscillations, when moving beyond the torus scale (Komossa \& Merritt 2008).

The kicking mechanism of the rSMBH is due to an anisotropic radiation of the gravitational wave in merging black holes and
detection of the rSMBHs will be an indirect observational evidence of the existence of the gravitational wave.
Despite of these astrophysical importance, none of the rSMBHs has been positively identified yet.
Thus, we have undertaken spectral decomposition project for the nearby SDSS QSOs to identify rSMBHs.

The paper is divided into four sections. 
In Section 2, the sample selection process of the rSMBH candidates is described. 
The results and discussion are presented in Sections 3, 
and a summary of the paper is presented in Section 4. 
Throughout this paper, the cosmology H$_0$ = 70 km s$^{−1}$ Mpc$^{−1}$, $\Omega_M$ = 0.3, and $\Omega_\Lambda$ = 0.7 are adopted.
 
\section{Selection Process of the rSMBH Candidates}

To select rSMBH candidates, we used a complete sample of 1271 z $<$ 0.25 SDSS DR7 QSO (Schneider et al. 2010, hereafter SDSS QSO) spectra.
The size $r_k$ of the rSMBHs is given by
\begin{equation}
r_{\rm k} = {GM_{\rm BH}\over{v_{\rm k}^2}} \approx 0.4
 \left({M_{\rm BH}\over{10^8\ {\rm M}_\odot}}\right)
\left({v_{\rm k}\over{10^3\ {\rm km\ s}^{-1}}}\right)^{-2} ~{\rm pc},
\end{equation}
where $M_{\rm BH}$ is mass of the black hole and $v_{\rm k}$ is the kick velocity (Merritt et al. 2009).
With typical values of $M_{\rm BH}=5\times 10^7 M_\odot$ and $v_{\rm k}$=200 $\kms$, the size of rSMBH becomes $\sim$5 pc, which is large enough to carry BLR.
Thus, if the rSMBHs are recoiled from the host galaxy, broad emission lines will show velocity offsets with respect to systemic velocity.
Fig. 1 illustrates model spectra of H$\alpha +$ [N II] for an object with no velocity offset in broad-line (Fig. 1a),
a BLR coming toward us (blueshifted 200 km$^{-1}$ with respect to systemic, Fig. 1b), 
and a BLR moving away from us (redshifted 200 km$^{-1}$ with respect to systemic, Fig. 1c).

As a first step for searching for rSMBH candidates, 
we have visually inspected the SDSS QSO spectra and selected spectra showing blueshifted broad-line with respect to systemic as shown in Fig. 1b.
We favor blueshifts because it will be easier to see the rSMBHs coming toward us rather than moving away from us due to extinction in the host galaxy.
The broad-line velocity offsets can be produced by 
gas motion in the accretion disk around a single BH (so called disk-emitters; Chen et al. 1989; Eracleous \& Halpern 1994; Chornock et al. 2010) or
binary SMBHs (Komossa et al. 2003; Zhou et al. 2004; Eracleous et al. 2012; Shen et al. 2013).
Thus, we have excluded spectra having widely separated and highly asymmetric two broad-line profiles (disk-emitters)
and double-peaked emission lines (binary SMBHs) in this stage.
For the systemic velocity, we used low-ionization (ionization potential = 10.36 eV) forbidden lines [S II]$\lambda 6716,6731$ which are expected to be less influenced by central engine.
If the [S II]$\lambda 6716,6731$ lines are not available, we used either [N II] $\lambda 6583$ or core component of [O III] $\lambda 5007$ lines since [O III] line often shows blueshifted wing component (i.e. Boroson 2005).
This step resulted in 124 candidates and for these objects, we have performed a detailed spectral decomposition.

In the spectral decomposition, we used the H${\alpha}$ line instead of the H${\beta}$ line since the H${\alpha}$ line is 
about 3 times stronger than the H${\beta}$ line and free from Fe II line contamination.
We used the IRAF/Specfit package with 3 component fits: i) power-law continuum, ii) broad emission line, and iii) narrow emission line.
We applied a single Lorentzian or a single Gaussian profile for the H${\alpha}$ broad-line component and a single Gaussian profile for the H${\alpha}$ and [N II] narrow line components.
The same Gaussian line widths were used for the narrow emission lines of H${\alpha}$ and [N II], and
a fixed value of $1/3$ was used for the [NII]6548 to [NII]6583 line ratio.
From spectral decomposition, we have identified 37 rSMBH candidates whose H${\alpha}$ broad-lines are blueshifted
more than 1.5\AA\, which corresponds to instrumental dispersion of 69 km s$^{\rm -1}$ of the SDSS spectra, with respect to systemic velocity.
In this stage, we applied one more constraint before selecting the final rSMBH candidates.
It is known that line widths of H$\alpha$ and H$\beta$ broad-lines are well correlated (i.e. Eq. 3 of Greene \& Ho 2005).
This suggests that their line emitting gases are co-spatial.
Thus, if the H$\alpha$ broad-line velocity offset was produced by recoiling BLR, 
we should see a similar amount of velocity offset in H$\beta$ broad-line too.
We have measured velocity offsets in the H$\beta$ broad-line using 4 components fit:
i) power-law continuum, ii) broad emission line, iii) narrow emission line, and iv) [Fe II] lines.
To select the final rSBMH candidates, we have applied the following arbitary selection criterion: 
we excluded sources whose H${\beta}$ broad-line velocity offset is either
50\% smaller or 50\% larger than that of the H${\alpha}$ broad-line velocity offset.
After applying this criterion, the final number of rSMBH candidates becomes 26.
Fig. 2 shows correlation of velocity offset between the H${\alpha}$ and H${\beta}$ broad-lines,
\textbf{where solid and dotted lines are one-to-one correspondence and $\pm 50\%$ deviation from the one-to-one correspondence, respectively.
For the final 26 rSMBH candidates,} we see a good correlation of velocity offsets between H${\alpha}$ and H${\beta}$ broad-lines ($r=0.90$).
The H${\beta}$ broad-lines are on average $32\pm 38$ \kms more blueshifted than the H${\alpha}$ broad-lines.

Fig. 3 shows results of spectral decompositions for H${\alpha}$ and H${\beta}$ lines for the final rSMBH candidates, 
where black, cyan, blue, green, magenta, and red lines represent 
data, power-law continuum, broad emission line, narrow emission lines, [O III] lines, and model (sum of all fitting components), respectively.
Dotted lines represent [Fe II] lines and vertical dotted line in each plot represents rest frame line center for the H${\alpha}$ and H${\beta}$ lines.
Fitting residuals (data/model in percentage) are shown on the bottom of each plot.
As mentioned earlier, we used Gaussian or Lorentzian profile to fit H${\alpha}$ broad-line,
but interestingly the H${\alpha}$ broad-line profiles for the rSMBH candidates are all found to be Lorentzian 
which suggests they originate by a turbulent velocity field.
Carmona-Loaiza et al. (2015) find that turbulence dominates corotating inflows near the SMBHs and significantly enhance accretion activity.
Thus, turbulence could be a dominant velocity field in the accretion disk of the rSMBHs.

\section{Results and Discussion}

\subsection{Properties of rSMBH candidates}

Table 1 lists basic properties of the final 26 rSMBH candidates.
The fraction of the rSMBH candidates in SDSS QSO spectra with z $<$ 0.25 is $\sim$2\%.
If we assume the fraction of the rSMBH moving away from us has a similar fraction, about 4\% of the SDSS QSOs are rSMBHs.
\textbf{A histogram of projected recoil velocities measured from H${\alpha}$ broad-line is presented in Fig. 4a.
The projected recoil velocities range from $-76\ \kms$ to $-307\ \kms$ with an average of $-149\pm58\ \kms$. 
Except for one, all candidates have recoil velocities smaller than absolute value of $-220\ \kms$.
The lack of detecting rSMBH candidates with large recoil velocities could be explained either by a finite size of SDSS fiber radius (see \S 3.2) or orbital configuration in merging black holes that produces large recoil velocities ($<-300\ \kms$) is rare.
Fig. 4b shows a plot of recoil velocity vs. velocity dispersion of the rSMBH candidates. 
Though the correlation is poor (r=0.17), we see a trend that the broad-line velocity offset 
tends to increase with decreasing velocity dispersion. 
Velocity dispersion has a linear correlation with mass (i.e. M-$\sigma$ relation).
Thus, the rSMBHs with large recoil velocities are more likely to be found in less massive galaxies
and the rSMBHs with small recoil velocities are more likely to be found in more massive galaxies.
This can naturally be understood in terms of escape velocity since massive galaxies have large escape velocities
and exert more gravitational drag to the rSMBHs and trap them to undergo oscillatory motion.
Infrared luminosity vs. velocity offset plot is presented in Fig. 4c where we see only a weak correlation (r=0.31).
}

\textbf{
The infrared luminosities for our sample range from 10$^{10.69}$ to 10$^{12.11}$ L$_\odot$ and 
the corresponding range of star formation rates (SFR) estimated from Kennicutt \& Evans (2012, i.e. SFR/M$_\odot$yr$^{-1}$ = 1.5 $\times 10^{-10}$ L$_{FIR}$/L$_\odot$) are 7 to 190 M$_\odot$yr$^{-1}$.
Predicted SFRs in the center of SMBH-displaced host galaxies range from a few M$_\odot$yr$^{-1}$ (Blecha et al. 2011) 
to a few $\times 10^3$ M$_\odot$yr$^{-1}$ (Sijacki et al. 2011). 
Our estimated value fits in the middle of these two predictions.
}

The SDSS g, r, and i composite images, which are deep enough ($\sim26$ mag arcsec$^{-2}$, Kniazev et al. 2004) to detect tidal features, of the rSMBH candidates are presented in Fig. 5. 
We find that one object shows strong tidal tails and is clearly interacting (J075819.69+421935.1), 
four objects show short tidal tails (J080742.46+375332.1, J115341.16+101754.0, J130103.5+065711.7, and J155504.78+134335.8), 
one object shows an off-center nucleus (J091617.53+303037.9), and
five objects have nearby companions (J085946.35+274534.8, J105752.69+105037.9, 
J115632.23+112653.8, J145824.46+363119.5, and J163734.52+305403.5).
For these latter five sources, we do not see obvious signs of tidal interaction between main galaxy and its companion.
The fraction of tidal features in our candidates is about 23\% (6 out of 26).

\subsection{Implications of the recoils}

From the projected recoil velocities, we can estimate projected spatial offsets between the rSMBHs and center of host galaxy.
The projected recoil velocities are the radial components of the actual ones and 
the projected spatial offsets are the transversal components of the actual ones.
If we assume random recoil direction (polar angle $\theta$ = azimuthal angle $\phi$ = 45$^\circ$) for ease of calculation,
the expected spatial offset will be the recoil velocity times the AGN lifetime.
Column 8 in Table 1 lists the expected spatial offsets calculated by adopting an AGN lifetime of 10 Myr for rSMBHs (Blecha et al. 2011).
The estimated spatial offsets are maximum values for the AGN lifetime of 10 Myr and range from 0.21\as \ to 1.97\as \ (0.79 kpc to 3.14 kpc) with a mean of 0.51\as$\pm$0.35\as \ (1.52$\pm$0.59 kpc).
The adopted AGN lifetime of 10 Myr seems to be reasonable since all but one of the estimated separations fall within the SDSS fiber radius of 1.5\as.
If the estimated spatial offsets were larger than 1.5\as, it would not been observed with the SDSS spectrograph with 1.5\as\ radius fiber.
If the SMBH is recoiled at z=0.2 to a random direction with recoil velocity of $\sim 470$\kms, it will move outside of the SDSS fiber radius in 10 Myr.
The limited radius of SDSS fiber could explain why we couldn't detect high-velocity rSMBHs even if they exist.

The recoil velocities we have measured are only about 10\% to 20\% of the typical escape velocity in elliptical galaxy ($v_{\rm e}\approx$1500 km s$^{-1}$).
In such a case, the rSMBH candidates will undergo damped oscillations around the center of host galaxy (Gualandris \& Merritt 2008; Blecha et al. 2011).
Blecha et al. (2011) predicted that the oscillating SMBHs may be up to five times less massive than their
stationary counterparts and could be a source of intrinsic scatter in the SMBH and stellar bulge mass scaling law.
We have investigated this possibility from correlations of black hole mass (M$_{BH}$) vs. velocity dispersion ($\sigma_v$), and
M$_{BH}$ vs. Eddington ratio (L$_{bol}$/L$_{Edd}$) for SDSS QSOs and rSMBH candidates.
The velocity dispersion $\sigma_v$ was estimated from line widths of [S II]6716,6731 (FWHM ([S II])/2.35; i.e. Komossa \& Xu (2007))
and the bolometric luminosity L$_{bol}$ was adopted from Shen et al. (2011).
The black hole mass M$_{BH}$ was calculated from virial mass formula that depends on the H$\alpha$ line 
(Greene \& Ho 2005; Reines et al. 2013):
\begin{eqnarray}
{\rm M_{BH}} & = & f \times 10^{6.57} \left(\frac{\rm L_{H\alpha}}{10^{42}~{\rm erg~s^{-1}}} \right)^{0.47}
\left(\frac{\rm FWHM_{H\alpha}}{10^{3}~\mathrm{\kms}} \right)^{2.06} \msun ,
\end{eqnarray}
where $f$, L$_{H\alpha}$, and FWHM$_{H\alpha}$, are virial factor that matches measured virial
product into the actual black hole mass, H$\alpha$ broad-line luminosity and FWHM of H$\alpha$ broad-line, respectively.
For the virial factor $f$, we adopted the latest value of $f=1.12^{+0.36}_{-0.27}$ from Woo et al. (2015).

The correlation between the M$_{BH}$ and $\sigma_v$ is plotted in Fig. 6a, \textbf{where blue circles and green dots represent rSMBH candidates and SDSS QSOs, respectively. Solid lines represent least square fits for each data.
We find that M$_{BH}$ of the rSMBH candidates are on average 5.2$\pm$3.2 (median=4.1) times smaller than that in SDSS QSOs
and causes a large scatter in the M$_{BH}$ and $\sigma_v$ relation.
For a moderate kick velocity ($0 < v_{\rm k} < 0.9 v_{\rm e}$),
the mass of the rSMBH grows about 3 times more than initial mass (Sijacki et al. 2011).
Even if we assume our rSMBH candidates are in the initial stage of recoiling, the mass of full grown rSMBH is still 2 times smaller than their stationary counterpart. 
When stellar mass black holes are merging, only $\sim$5\% of total mass is carried away by a gravitational wave (Abbott et al. 2016).
Unlike stellar mass BH merging, a significant fraction of mass could be carried away by gravitational waves in the merging process of the SMBHs, which could be one of the possibilities of small SMBH mass of the rSMBHs.
}
Recent simulations suggest that the rSMBHs could be a source of scatter in M$_{BH}$ and host galaxy bulge mass scaling law (Volonteri 2007; Blecha et al. 2011).
The simulations predicted the black hole mass shortfall preferentially occur at high velocity recoil events ($v_{\rm k}\approx v_{\rm e}$ for Volonteri (2007) and $v_{\rm k} \geq 0.8\times v_{\rm e}$ for Blecha et al. (2011))
and the black hole mass for rapidly recoiling rSMBH can be up to 5 times smaller than their stationary counterpart (Blecha et al. 2011).
Note, however that the recoil velocities we have measure are much smaller than their kick velocities.
The kick velocity $v_{\rm k}$ is the initial velocity when the SMBH is recoiled at the center of host galaxy and
the recoil velocity we have measured is recoil velocity at spatially offset position.
To be compatible with the simulations, the kick velocity must have suffered a significant gravitational drag after it kicked out.

The correlation between the M$_{BH}$ and Eddington ratio $L_{bol}/L_{Edd}$ is shown in Fig. 6b where blue circles and green dots represent 
rSMBH candidates and SDSS QSOs, respectively.
The solid line is a least-square fit to the SDSS QSOs.
The Eddington ratio of all of the rSMBH candidates is larger than 0.1 with an average of 0.52$\pm$0.27 (median=0.48) 
whereas it is 0.19$\pm$0.28 (median=0.11) in SDSS QSOs.
The small black hole mass in Fig. 6a and large Eddington ratio in Fig. 6b suggest that rSMBHs are 
actively accreting mass and could be in the stage of growing SMBH mass until they can reach full-grown SMBHs.
If most of the interacting galaxies must go through this stage,
then it will be an important evolutionary path in the galaxy evolution we have missed so far.

It is suggested that star formation in galaxies is regulated by AGN feedback and outflow (Fabian 2012; Tombesi et al. 2015). 
If the central AGN is displaced by a recoil event, it will no longer result in quenching but instead enhancing central star formation in the host galaxy (Blecha et al. 2011; Sijacki et al. 2011).
If this is the case, the host galaxy of rSMBHs will contain a significant amount of gas and dust for star formation and could be very luminous in the infrared.
To calculate the infrared luminosity for each host galaxy, 
we used IRAS ADDSCAN/SCANPI (Helou et al. 1988).
For some sources, the default parameter values in ADDSCAN/SCANPI yielded a poor fit and could not be used to compute the IRAS fluxes.
In such a case, we adjusted the ADDSCAN/SCANPI source and background fitting ranges from default values 
(i.e. background fitting range 30 arcmin and source fitting ranges 3.2 arcmin (60 $\mu m$) and 6.4 arcmin (100 $\mu m$)) to appropriate ones.
The IRAS 60 $\mu m$ and 100 $\mu m$ fluxes and calculated infrared luminosities are listed in Table 1.
We find that most of them are infrared luminous galaxies; either LIRGs (L$_{FIR} > 10^{11}$ L$_{\odot}$) or ULIRGs (L$_{FIR} > 10^{12}$ L$_{\odot}$).
Furthermore, $\sim60\%$ of them (16 out of 26) show Wolf-Rayet features suggesting evidence of recent star formation activity in these galaxies.
The typical morphology of the LIRGs/ULIRGs are interacting/merging galaxies with tidal features (Sanders et al. 1988; Kim et al. 2013; Evans et al. 2016).
The fraction of tidal features in rSMBH candidates (23\%) is much smaller than that of the LIRGs/ULIRGs (80\%, Kim et al. 2013) suggesting
that, if the AGN activity was triggered by interactions, majority of the rSMBH candidates are advanced mergers.
If this is the case, gravitational recoil event could happen long after the merging event when the tital features were almost faded.

\subsection{Do broad-line offsets in rSMBH candidates really originated from recoiling SMBHs?}

\subsubsection{Possible origin of broad-line velocity offset}

The broad-line velocity offsets can be produced by disk-emitters or binary SMBHs, 
but in the candidate selection process, we have excluded these sources.
 
An AGN outflow powered by radiation pressure from the accretion disk or radio jets can produce 
a blueshifted broad-line velocity offset (Harrison et al. 2012; Genzel et al. 2014).
It is suggested that the AGN outflows can also produce blue wings in the [O III]$\lambda \lambda$4959,5007 emission lines (Heckman et al. 1981, Whittle et al. 1988, Crenshaw et al. 2010).
Thus, if the blueshifted H${\alpha}$ broad emission line is produced by outflow,
we would expect to observe a correlation of the velocity offsets between the H${\alpha}$ broad-line and [O III]$\lambda$5007 wing component.
To measure the velocity offset in the [O III] wing component,
we have decomposed the [O III]$\lambda$5007 line into core and wing components for all of the rSMBH candidates.
The fitting results are plotted in Fig. 3 and listed in Column 2 of Table 2.
The correlation between the velocity offsets in H${\alpha}$ broad-line and [O III]$\lambda$5007 wing component is
plotted in Fig. 7,
where dashed and solid lines represent one to one relation and least-square fit of the data, respectively.
We find only a weak correlation (r=0.35) of velocity offsets between the H${\alpha}$ broad-line and [O III]$\lambda$5007 wing component.
This suggests that the AGN outflow could be responsible for the broad-line offset for a few individual sources, but not likely for the whole sample.

Another scenario for producing narrow lines plus blueshifted broad lines is an interacting/merging system with double nuclei: 
one nucleus is Seyfert producing broad and narrow lines and the other one is either LINER or H II galaxy producing narrow lines.
As discussed in Section 3.1, only 6 rSMBH candidates show tidal signatures.
If these rSMBHs are Seyfert and LINER/HII pairs, 
we would observe two distinct sets of narrow forbidden lines in the spectra.
We have inspected the spectra of all 26 rSMBH candidates, but none of them have two sets of narrow lines.
Thus, it is unlikely that the broad-line velocity offsets in rSMBH candidates are caused by interacting/merging galaxies.

\subsubsection{High-excitation coronal lines}

The presence of high-excitation forbidden lines whose ionization potential $I_p$ is larger than He II ionization edge (54.4 eV)
is an unambiguous sign of AGN activity.
It is known that the high-excitation coronal lines are often blueshifted by a few hundreds km s$^{-1}$ with respect to systemic
(Appenzeller \& Oestreicher 1988; Erkens et al. 1997; Mullaney et al. 2009; Gelbord et al. 2009).
If the rSMBHs are indeed recoiling, the blueshift of the high-excitation coronal lines will be the sum of 
the recoil velocity plus the blueshift component of the coronal lines (i.e. more blueshifted than their stationary counterpart).
Thus, the high-excitation coronal lines can be used to probe the origin of the broad-line velocity offset in rSMBHs.
We have inspected spectra of the rSMBHs and find 23 out of 26 (88\%) and 11 out of 26 (42\%) of the rSMBHs
have high-excitation coronal lines of [Ne V]$\lambda$3426 ($I_p$=97.11 eV) and [Fe VII]$\lambda$6087 ($I_p$=99.00 eV), respectively.
We found that most of the coronal lines in our candidates have two components: a narrow core component and blueshifted broad wing component.
Spectral decomposition was performed for [Ne V]$\lambda$3426 and [Fe VII]$\lambda$6087 lines and 
their results are listed in Table 3.
As summarized in Table 3, the average velocity offsets for [Ne V]$\lambda$3426 ([Fe VII]$\lambda$6087) are 
$-$235$\pm$190 km s$^{-1}$ and $-$690$\pm$430 km s$^{-1}$ (-340$\pm$190 km s$^{-1}$ and $-$560$\pm$180 km s$^{-1}$) for core and wing component, respectively.
While studying ionization properties for nearby (z$<$0.4) 63 SDSS Seyfert galaxies, 
Gelbord et al. (2009) found velocity offsets of $-$74$\pm$16 km s$^{-1}$ in core component and $-$296$\pm$28 km s$^{-1}$ in wing component of the [Fe VII]$\lambda$6087 line.
Compared to their measurements, [Fe VII]$\lambda$6087 in rSMBH are on average about $-$270 km s$^{-1}$ and $-$260 km s$^{-1}$ more blueshifted for core and wing component, respectively.
Their sample can be regarded as stationary AGNs since their H$\alpha$ broad-line velocity offset with respect to systemic is only $-$3$\pm$26 km s$^{-1}$.

Thus, this result offers further evidence (although still circumstancial) that rSMBH candidates are indeed recoiling with a mean recoil velocity of about $-$265 km s$^{-1}$.
For a combined sample of [Ne V]$\lambda$3426 and [Fe VII]$\lambda$6087, the average velocity offsets become 
$-$270$\pm$190 km s$^{-1}$ and $-$665$\pm$400 km s$^{-1}$ for core and wing component, respectively.
Compared to the core component, the $\sim -$400 km s$^{-1}$ larger velocity offset in wing component could indicate that these lines originate closer to the AGN center.

Velocity offset plots between H$\alpha$ broad-line and coronal lines are presented in
Fig. 8a (core component) and Fig. 8b (wing component)
where blue and red circles represent velocity offset for [Ne V]$\lambda$3426 and [Fe VII]$\lambda$6087, respectively.
The dotted lines represent one-to-one correspondence and the dashed lines are least-square fit of the data.
We find a weak correlation ($r$=0.26) between the H$\alpha$ broad-line and core components of the coronal lines,
and no correlation ($r$=0.07) between the H$\alpha$ broad-line and wing components of the coronal lines.
Though they have poor correlations, best fit slopes of the dashed lines are almost parallel to those of the dotted lines (i.e. coronal lines have constant velocity offsets with respect to H$\alpha$ broad-line).
These two results suggest that the velocity offset between H$\alpha$ broad-line and coronal lines are 
not locally correlated, probably because the physical/kinematic conditions of the line emitting regions are different,
but globally well correlated due to recoil motion.
Further evidence that the rSMBH candidates are actually recoiling is found in the velocity offset between H$\alpha$ broad-line and coronal lines.
As shown in Fig. 8c, we find a good correlation (r=0.77) of velocity offset between two coronal lines [Ne V]$\lambda$3426 and [Fe VII]$\lambda$6087.
This, along with a similar $I_p$ for [Ne V]$\lambda$3426 (97.11 eV) and [Fe VII]$\lambda$6087 (99.00 eV), indicate that these lines are originated from kinematically/spatially similar regions.

\subsubsection{Diagnostic line ratios}

Diagnostic line ratios (Baldwin, Phillips \& Terlevich 1981; Veilleux \& Osterbrock 1987; Kewley et al. 2006) of the rSMBH candidates have been 
investigated to see if they can provide a clue for the origin of broad-line velocity offset.
Fig. 9 shows narrow line ratio plots of [N II]/H$\alpha$ vs. [O III]/H$\beta$ (Fig. 9a) and [S II]/H$\alpha$ vs. [O III]/H$\beta$ (Fig. 9b),
where blue circles and green dots represent rSMBH candidates and SDSS QSOs, respectively.
The overall distribution of line ratios for the rSMBHs are similar to those of the SDSS QSOs. 
However, we notice that a higher fraction of H II region sources is found in rSMBH candidates than in SDSS QSOs:
about 19\% and 23\% of the rSMBHs fall in H II region in [N II]/H$\alpha$ and [S II]/H$\alpha$ plots, respectively 
whereas only 3\% and 11\% of the SDSS QSOs fall in H II region in [N II]/H$\alpha$ and [S II]/H$\alpha$ plots, respectively.
The large fraction of H II region sources in rSMBH candidates 
could be related with a combined effect of spatially offset AGN and density-stratified NLR.

To investigate further, we have divided the rSMBH candidates into two groups: 
H II region group (5 sources inside red-dotted circles in the plots) and AGN group (the rest of the sources).
We find that narrow line luminosities in H II region rSMBH candidates
are on average about 4.0, 5.2, and 6.5 times smaller for [O III], [N II], and [S II] lines, respectively than those in AGN rSMBH candidates.
If the H II rSMBH are spatially more offset from center of host galaxy than the AGN rSMBHs, 
they will have smaller electron density and their narrow line luminosities will be weaker than that in AGN rSMBHs under density-stratified NLR environment (e.g., Filippenko \& Halpern 1984; Veilleux 1991).
We tested this hypothesis by calculating electron density $N_e$ from [S II]6716/[S II]6731 ratio using
the Temden task in Nebular package of the IRAF/STSDAS (Shaw \& Dufour 1995)
assuming typical temperature of $T=1.0\times 10^4$K in line emitting region of the AGNs (Osterbrock \& Ferland 2006).
The Temden task numerically solves the equilibrium equation for a 5-level atom approximation developed by De Robertis et al. (1987).
As expected, the mean $N_e$ in the H II region rSMBH candidates ($22^{+32}_{-0}$ cm$^{-3}$, median=0 cm$^{-3}$) is
about 10 times smaller than that in AGN rSMBH candidates ($216^{+218}_{-152}$ cm$^{-3}$, median=266 cm$^{-3}$). 
In Section 3.1, we have estimated the expected spatial offset for the rSMBH by assuming random recoil direction.
Even if we used this crude assumption,
the spatial offset in H II region rSMBH candidates (mean=$2.1\pm 1.0$ kpc, median=1.7 kpc) is
$\sim$30\% larger than that in AGN rSMBH candidates (mean=$1.4\pm 0.5$ kpc, median=1.3 kpc). 

The same argument can be applied to the 26 rSMBH candidates versus the SDSS QSOs.
If the rSMBH candidates are spatially offset, their $N_e$ will be smaller than that of the SDSS QSOs.
The calculated mean $N_e$ are 172$^{+209}_{-145}$ cm$^{-3}$ (median=165 cm$^{-3}$) and 480$^{+650}_{-320}$ cm$^{-3}$ (median=435 cm$^{-3}$)
for the rSMBH candidates and SDSS QSOs, respectively.
Indeed, the mean $N_e$ in rSMBH candidates is only $\sim$1/3 of that in SDSS QSOs.
This could be another evidence that the rSMBH candidates are actually recoiling.
 
\section{Summary}

We have performed a spectral decomposition to search for recoiling supermassive black hole (rSMBH) candidates in the SDSS QSOs with $z<0.25$.
We reported on the properties of the rSMBH candidates and on the mounting evidence that these candidates are in fact recoiling.

$\bullet$ 
Out of 1271 QSOs, we have identified 26 rSMBH candidates that are recoiling toward us.
If we assume a similar number of rSMBHs are recoiling to opposite direction, then $\sim$4\% of the SDSS QSOs with $z<0.25$ are rSMBHs.

$\bullet$
The projected recoil velocities of the rSMBH candidates range from $-76\ \kms$ to $-307\ \kms$ with an average of $-149\pm58\ \kms$.
The expected spatial offsets range from 0.21\as \ to 1.97\as \ (0.79 kpc to 3.14 kpc) with a mean of 0.51\as$\pm$0.35\as \ (1.52$\pm$0.59 kpc)
if we assume a random recoil direction and an AGN lifetime of 10 Myr for the rSMBH candidates. 

\textbf{
$\bullet$
The rSMBH candidates in massive hosts are more likely to undergo oscillatory motion due to large escape velocity of the host galaxy.}

$\bullet$
The line profile of the H$\alpha$ broad-line in all rSMBH candidates was found to be Lorentzian and suggests turbulence is a dominant velocity field in the accretion disk of the rSMBHs.

$\bullet$
We find that the black hole mass $M_{BH}$ of the rSMBH candidates is on average 5.2$\pm$3.2 (median=4.1) times smaller than 
that of their stationary counterparts and causes a large scatter in $M_{BH}-\sigma_*$ relation.
\textbf{Unlike stellar mass BH merging, a significant fraction of mass could be
carried away by gravitational wave in the merging process of the SMBHs, which
could be one of the possibilities of small SMBH mass of the rSMBHs.}
The Eddington ratio of all of the rSMBH candidates is larger than 0.1 with average of 0.52$\pm$0.27 (median=0.48)
suggesting they are actively accreting mass.

$\bullet$
Most of the rSMBH candidates are found to be LIRGs or ULIRGs, 
but unlike typical LIRGs/ULIRGs only 23\% shows a sign of tidal features suggesting majority of them are advanced mergers. 

$\bullet$
The possibility that AGN outflows or double nuclei (Seyfert $+$ LINER/HII) cause the broad-line velocity offsets were examined,
but we could not find strong evidences, except that the broad-line velocity offset in a few individual sources could have been originated by AGN outflow.

$\bullet$
Slopes in the velocity offset plots of H$\alpha$ broad-line vs. high-excitation coronal lines suggest that the rSMBH candidates are actually recoiling.
Comparison of [Fe VII]$\lambda$6087 velocity offsets between rSMBH candidates and stationary AGNs suggest that the rSMBH candidates are recoiling on average velocity of $-$265 km s$^{-1}$.

$\bullet$
We find that electron density in the NLR of the H II rSMBHs is about 1/10 of that in AGN rSMBHs.
This can be explained if the H II rSMBHs are more spatially offset than the AGN rSMBHs under density-stratified NLR environment.
The mean electron density in rSMBH candidates is found to be 1/3 of that in SDSS QSOs suggesting the former was more spatially offset than the latter. This can be another evidence that the rSMBH candidates are actually recoiling.

As summarized above, there is much evidence to suggest that broad-line velocity offsets in rSMBH candidates are originated from recoiling SMBH.
To confirm this, we need evidence for spatial offset between the rSMBH and center of host galaxy. 
We have submitted adaptive optics imaging proposal to identify and measure spatial offset
and integral-field spectroscopy to study velocity field in the rSMBH candidates.
If confirmed, the result will have significant implications in astrophysics including 
formation and evolution of SMBHs, black hole binary dynamics, SMBH and bulge mass scaling law, and AGN unification theories.

The authors thank the anonymous referee for comments and suggestions that greatly improved this paper.
We also thank Y. Shen for useful discussions.
This research has made use of the NASA/IPAC Extragalactic Database
(NED) which is operated by the Jet Propulsion Laboratory, California
Institute of Technology, under contract with the National Aeronautics and Space Administration. 
D.C.K., A.S.E., and S.S acknowledge support from the National Radio Astronomy Observatory (NRAO) and G.C.P. was supported by a FONDECYT Postdoctoral Fellowship (No.\ 3150361). 
The National Radio Astronomy Observatory is a facility of the National Science Foundation operated under cooperative agreement by Associated Universities, Inc.

\clearpage

\begin{deluxetable}{lccccccccccc}
\tabletypesize{\scriptsize}
\tablewidth{0pt}
\tablecaption{Properties of the Recoiling SMBH Candidates}
\tablehead{
\multicolumn{1}{c}{Name} &
\multicolumn{1}{c}{z} &
\multicolumn{1}{c}{r} &
\multicolumn{1}{c}{f60} &
\multicolumn{1}{c}{f100} &
\multicolumn{1}{c}{${L_{FIR} \over L_\odot}$} &
\multicolumn{1}{c}{$\Delta V$(H$\alpha$)} &
\multicolumn{1}{c}{Spatial offset} &
\multicolumn{1}{c}{$\Delta V$(H$\beta$)} &
\multicolumn{1}{c}{$\sigma_v$} &
\multicolumn{1}{c}{${M_{BH} \over M_\odot}$} &
\multicolumn{1}{c}{${L_{bol} \over L_{Edd}}$} \\
\multicolumn{1}{c}{SDSS J} &
\multicolumn{1}{c}{} &
\multicolumn{1}{c}{mag} &
\multicolumn{1}{c}{Jy} &
\multicolumn{1}{c}{Jy} &
\multicolumn{1}{c}{log} &
\multicolumn{1}{c}{\kms} &
\multicolumn{1}{c}{\arcsec\ \ (kpc)} &
\multicolumn{1}{c}{\kms} &
\multicolumn{1}{c}{\kms} &
\multicolumn{1}{c}{log} &
\multicolumn{1}{c}{}\\
\multicolumn{1}{c}{(1)} &
\multicolumn{1}{c}{(2)} &
\multicolumn{1}{c}{(3)} &
\multicolumn{1}{c}{(4)} &
\multicolumn{1}{c}{(5)} &
\multicolumn{1}{c}{(6)} &
\multicolumn{1}{c}{(7)} &
\multicolumn{1}{c}{(8)} &
\multicolumn{1}{c}{(9)} &
\multicolumn{1}{c}{(10)} &
\multicolumn{1}{c}{(11)} &
\multicolumn{1}{c}{(12)}
}
\startdata
$075819.69+421935.1$ & 0.212 & 17.01 & 0.33 & 0.54 & 11.98 & $-154\pm 10$ &  0.46 (1.57) & $-226\pm 25$ & $ 156\pm 3 $ & 8.03 & $ 0.25^{+0.06}_{-0.08}$ \\
$080101.41+184840.7$ & 0.140 & 16.62 & 0.08 & 0.62 & 11.36 & $ -87\pm 10$ &  0.36 (0.89) & $-110\pm 4 $ & $ 150\pm 22$ & 7.38 & $ 0.65^{+0.16}_{-0.21}$ \\
$080742.46+375332.1$ & 0.230 & 18.18 & 0.06 & 0.15 & 11.40 & $ -76\pm 5 $ &  0.21 (0.78) & $-105\pm 10$ & $ 127\pm 2 $ & 7.40 & $ 0.45^{+0.11}_{-0.14}$ \\
$085946.35+274534.8$ & 0.245 & 17.71 & 0.15 & 0.26 & 11.79 & $ -85\pm 10$ &  0.23 (0.87) & $-105\pm 5 $ & $ 250\pm 17$ & 7.48 & $ 0.54^{+0.13}_{-0.17}$ \\
$090654.47+391455.3$ & 0.241 & 17.78 & 0.12 & 0.12 & 11.60 & $-137\pm 36$ &  0.37 (1.40) & $-101\pm 23$ & $ 188\pm 21$ & 7.41 & $ 0.74^{+0.18}_{-0.24}$ \\
$091617.53+303037.9$ & 0.215 & 18.40 & 0.10 & 0.17 & 11.48 & $-208\pm 17$ &  0.61 (2.12) & $-236\pm 22$ & $ 158\pm 6 $ & 7.63 & $ 0.23^{+0.06}_{-0.07}$ \\
$092247.03+512038.0$ & 0.161 & 17.39 & 0.16 & 0.31 & 11.43 & $-212\pm 13$ &  0.78 (2.17) & $-215\pm 13$ & $  92\pm 13$ & 7.04 & $ 0.96^{+0.23}_{-0.31}$ \\
$105752.69+105037.9$ & 0.221 & 17.78 &\nodata&\nodata&\nodata&$ -91\pm 0 $&  0.26 (0.93) & $-137\pm 13$ & $ 148\pm 2 $ & 7.17 & $ 0.98^{+0.24}_{-0.31}$ \\
$115341.16+101754.0$ & 0.161 & 16.37 & 0.25 & 0.38 & 11.58 & $-202\pm 5 $ &  0.74 (2.07) & $-189\pm 25$ & $ 180\pm 3 $ & 7.91 & $ 0.27^{+0.07}_{-0.09}$ \\
$115632.23+112653.8$ & 0.226 & 17.52 & 0.15 & 0.42 & 11.81 & $-161\pm 6 $ &  0.45 (1.64) & $-191\pm 14$ & $ 157\pm 11$ & 7.53 & $ 0.51^{+0.12}_{-0.16}$ \\
$122749.13+321458.9$ & 0.137 & 17.66 & 0.16 & 0.42 & 11.34 & $ -77\pm 4 $ &  0.33 (0.79) & $ -91\pm 23$ & $ 172\pm 3 $ & 7.26 & $ 0.23^{+0.06}_{-0.07}$ \\
$130103.50+065711.7$ & 0.233 & 17.77 & 0.07 & 0.10 & 11.38 & $-208\pm 11$ &  0.57 (2.13) & $-290\pm 23$ & $ 153\pm 7 $ & 7.88 & $ 0.17^{+0.04}_{-0.05}$ \\
$132018.19+390722.7$ & 0.236 & 18.35 & 0.11 & 0.19 & 11.62 & $-132\pm 26$ &  0.36 (1.34) & $-165\pm 54$ & $ 222\pm 3 $ & 7.64 & $ 0.21^{+0.05}_{-0.07}$ \\
$132820.40+240927.4$ & 0.223 & 17.44 & 0.04 & 0.07 & 11.13 & $-136\pm 15$ &  0.39 (1.39) & $-118\pm 10$ & $ 179\pm 3 $ & 7.59 & $ 0.65^{+0.16}_{-0.21}$ \\
$134537.81+633130.5$ & 0.176 & 17.90 & 0.05 & 0.11 & 11.03 & $-110\pm 0 $ &  0.38 (1.12) & $-128\pm 10$ & $ 126\pm 15$ & 7.19 & $ 0.52^{+0.12}_{-0.16}$ \\
$134640.03-030013.1$ & 0.224 & 18.32 & 0.06 & 0.11 & 11.31 & $-202\pm 10$ &  0.57 (2.06) & $-254\pm 15$ & $ 115\pm 10$ & 7.18 & $ 0.71^{+0.17}_{-0.23}$ \\
$135516.56+561244.6$ & 0.122 & 16.77 & 0.07 & 0.06 & 10.69 & $-153\pm 5 $ &  0.71 (1.56) & $-103\pm 18$ & $ 205\pm 2 $ & 7.05 & $ 0.97^{+0.23}_{-0.31}$ \\
$140407.87+185208.3$ & 0.245 & 18.13 & 0.18 & 0.42 & 11.93 & $-107\pm 11$ &  0.28 (1.09) & $-138\pm 0 $ & $ 135\pm 5 $ & 7.51 & $ 0.43^{+0.10}_{-0.14}$ \\
$145824.46+363119.5$ & 0.247 & 16.96 & 0.10 & 0.22 & 11.67 & $-185\pm 12$ &  0.49 (1.89) & $-275\pm 12$ & $ 130\pm 3 $ & 7.72 & $ 0.72^{+0.17}_{-0.23}$ \\
$145840.41+195218.6$ & 0.223 & 17.68 & 0.20 & 1.16 & 12.11 & $-127\pm 7 $ &  0.36 (1.30) & $-179\pm 9 $ & $ 150\pm 11$ & 7.46 & $ 0.37^{+0.09}_{-0.12}$ \\
$155318.72+170202.9$ & 0.162 & 17.62 & 0.14 & 0.09 & 11.23 & $-217\pm 0 $ &  0.80 (2.22) & $-322\pm 29$ & $ 138\pm 3 $ & 6.96 & $ 1.00^{+0.24}_{-0.32}$ \\
$155504.79+134335.8$ & 0.174 & 17.49 & 0.11 & 0.27 & 11.39 & $ -84\pm 6 $ &  0.29 (0.86) & $-120\pm 0 $ & $ 140\pm 8 $ & 7.72 & $ 0.26^{+0.06}_{-0.08}$ \\
$160216.00+111926.9$ & 0.156 & 17.15 & 0.26 & 1.52 & 11.88 & $-160\pm 5 $ &  0.61 (1.64) & $-227\pm 15$ & $ 127\pm 12$ & 7.59 & $ 0.37^{+0.09}_{-0.12}$ \\
$163734.52+305403.5$ & 0.241 & 18.41 & 0.07 & 0.07 & 11.36 & $-175\pm 25$ &  0.47 (1.79) & $-211\pm 44$ & $ 157\pm 3 $ & 7.18 & $ 0.72^{+0.17}_{-0.23}$ \\
$171304.46+352333.5$ & 0.085 & 16.20 & 0.14 & 0.49 & 10.90 & $-307\pm 8 $ &  1.97 (3.14) & $-388\pm 25$ & $ 158\pm 9 $ & 7.13 & $ 0.38^{+0.09}_{-0.12}$ \\
$205418.80+004915.9$ & 0.228 & 18.05 & 0.17 & 0.72 & 11.97 & $ -80\pm 6 $ &  0.23 (0.82) & $ -87\pm 12$ & $ 136\pm 8 $ & 7.58 & $ 0.35^{+0.08}_{-0.11}$ \\
\enddata

\tablenotetext{\ } {{\it Col 1:}\ Object name.}
\tablenotetext{\ } {{\it Col 2:}\ Redshift.}
\tablenotetext{\ } {{\it Col 3:}\ SDSS r magnitude.}
\tablenotetext{\ } {{\it Col 4:}\ IRAS ADDSCAN/SCANPI 60 $\mu$m flux.}
\tablenotetext{\ } {{\it Col 5:}\ IRAS ADDSCAN/SCANPI 100 $\mu$m flux.}
\tablenotetext{\ } {{\it Col 6:}\ Far-Infrared luminosity.}
\tablenotetext{\ } {{\it Col 7:}\ Line of sight velocity offset from H$\alpha$ broad-line.}
\tablenotetext{\ } {{\it Col 8:}\ Estimated spatial offsets in arcsec and kpc units assuming AGN lifetime of 10 Myr.}
\tablenotetext{\ } {{\it Col 9:}\ Line of sight velocity offset from H$\beta$ broad-line.}
\tablenotetext{\ } {{\it Col 10:}\ Velocity dispersion.}
\tablenotetext{\ } {{\it Col 11:}\ Black hole masses estimated from virial method.}
\tablenotetext{\ } {{\it Col 12:}\ Eddington ratios calculated from virial black hole masses.}

\end{deluxetable}

\begin{deluxetable}{lcccccrrrr}
\tabletypesize{\scriptsize}
\tablewidth{0pt}
\tablecaption{Velocity offsets, diagnostic line ratios and electron density in the Recoiling SMBH}
\tablehead{
\multicolumn{1}{c}{Name} &
\multicolumn{1}{c}{[OIII]$_{w}$} &
\multicolumn{1}{c}{[NeV]$_{c}$} &
\multicolumn{1}{c}{[NeV]$_{w}$} &
\multicolumn{1}{c}{[FeVII]$_{c}$} &
\multicolumn{1}{c}{[FeVII]$_{w}$} &
\multicolumn{1}{c}{${\rm [O III] \over H\beta}$} &
\multicolumn{1}{c}{${\rm [N II] \over H\alpha}$} &
\multicolumn{1}{c}{${\rm [S II]\over H\alpha}$} &
\multicolumn{1}{c}{N$_e$} \\
\multicolumn{1}{c}{SDSS J} &
\multicolumn{1}{c}{\kms} &
\multicolumn{1}{c}{\kms} &
\multicolumn{1}{c}{\kms} &
\multicolumn{1}{c}{\kms} &
\multicolumn{1}{c}{\kms} &
\multicolumn{1}{c}{log} &
\multicolumn{1}{c}{log} &
\multicolumn{1}{c}{log} &
\multicolumn{1}{c}{cm$^{-3}$} \\
\multicolumn{1}{c}{(1)} &
\multicolumn{1}{c}{(2)} &
\multicolumn{1}{c}{(3)} &
\multicolumn{1}{c}{(4)} &
\multicolumn{1}{c}{(5)} &
\multicolumn{1}{c}{(6)} &
\multicolumn{1}{c}{(7)} &
\multicolumn{1}{c}{(8)} &
\multicolumn{1}{c}{(9)} &
\multicolumn{1}{c}{(10)}
}
\startdata
$075819.69+421935.1$ &  $-444\pm 17 $ & $-465\pm 27 $ & $-1788\pm 95 $ & $-616\pm 102$ & \nodata       &  1.22 & $-0.20$ &$-0.39$  &  294 \\ 
$080101.41+184840.7$ &  $-458\pm 74 $ & $-391\pm 27 $ & $-1084\pm 246$ & \nodata       & \nodata       &  0.53 & $-0.43$ &$-0.95$  &  228 \\ 
$080742.46+375332.1$ &  $-195\pm 29 $ & $ -22\pm 14 $ & $ -304\pm 131$ & $   7\pm 76 $ & $-456\pm 122$ &  1.05 & $-1.07$ &$-0.85$  &  116 \\ 
$085946.35+274534.8$ &  $-571\pm 135$ & $ -83\pm 86 $ & $ -787\pm 3  $ & \nodata       & \nodata       &  0.70 & $-0.11$ &$-0.68$  &    * \\ 
$090654.47+391455.3$ &  $-408\pm 240$ & $-468\pm 78 $ & \nodata        & \nodata       & \nodata       &  0.06 & $-0.79$ &$-1.78$  &   71 \\ 
$091617.53+303037.9$ &  $-277\pm 20 $ & $-178\pm 14 $ & $ -177\pm 97 $ & \nodata       & \nodata       &  1.19 & $ 0.06$ &$-0.27$  &  292 \\ 
$092247.03+512038.0$ &  $-566\pm 43 $ & \nodata       & \nodata        & \nodata       & \nodata       &  0.28 & $-0.72$ &$-1.23$  &    * \\ 
$105752.69+105037.9$ &  $-346\pm 20 $ & $-199\pm 52 $ & $ -602\pm 90 $ & $-507\pm 114$ & \nodata       &  0.96 & $-0.24$ &$-0.55$  &    4 \\ 
$115341.16+101754.0$ &  $-679\pm 28 $ & $-278\pm 17 $ & $ -898\pm 91 $ & $-225\pm 29 $ & $-627\pm 26 $ &  1.30 & $ 0.14$ &$-0.31$  &  276 \\ 
$115632.23+112653.8$ &  $-470\pm 56 $ & $-209\pm 47 $ & \nodata        & \nodata       & \nodata       &  0.57 & $-0.21$ &$-0.47$  &  207 \\ 
$122749.13+321458.9$ &  $-398\pm 40 $ & $ -28\pm 26 $ & $ -219\pm 41 $ & $-323\pm 29 $ & \nodata       &  1.04 & $-0.65$ &$-0.65$  &   44 \\ 
$130103.50+065711.7$ &  $-345\pm 19 $ & $-243\pm 31 $ & $ -676\pm 112$ & $-408\pm 73 $ & \nodata       &  0.81 & $-0.32$ &$-0.41$  &    * \\ 
$132018.19+390722.7$ &  $-272\pm 15 $ & $ -19\pm 9  $ & $ -633\pm 34 $ & \nodata       & \nodata       &  1.35 & $-0.06$ &$-0.09$  &  318 \\ 
$132820.40+240927.4$ &  $-439\pm 111$ & $-205\pm 55 $ & $ -773\pm 216$ & $-228\pm 47 $ & $-781\pm 77 $ &  0.51 & $-0.96$ &$-1.48$  &   41 \\ 
$134537.81+633130.5$ &  $-343\pm 19 $ & $-218\pm 22 $ & $ -363\pm 286$ & $-198\pm 23 $ & \nodata       &  0.87 & $ 0.06$ &$-0.17$  &  266 \\ 
$134640.03-030013.1$ &  $-499\pm 62 $ & \nodata       & \nodata        & \nodata       & \nodata       &  0.93 & $-0.24$ &$-0.57$  &  113 \\ 
$135516.56+561244.6$ &  $-351\pm 31 $ & $ -66\pm 3  $ & $ -344\pm 3  $ & $-201\pm 17 $ & $-369\pm 69 $ &  0.96 & $ 0.14$ &$-0.32$  &  919 \\ 
$140407.87+185208.3$ &  $-186\pm 23 $ & $ -91\pm 14 $ & $ -891\pm 97 $ & \nodata       & \nodata       &  1.08 & $ 0.05$ &$-0.36$  &  399 \\ 
$145824.46+363119.5$ &  $-360\pm 28 $ & $ -87\pm 17 $ & \nodata        & \nodata       & \nodata       &  0.84 & $-0.38$ &$-0.75$  &  607 \\ 
$145840.41+195218.6$ &  $-360\pm 51 $ & $-378\pm 42 $ & $ -361\pm 38 $ & $-496\pm 40 $ & \nodata       &  0.83 & $-0.03$ &$-0.52$  &  309 \\ 
$155318.72+170202.9$ &  $-363\pm 35 $ & $-180\pm 49 $ & $ -573\pm 129$ & $-520\pm 157$ & \nodata       &  0.80 & $-0.12$ &$-0.53$  &  127 \\ 
$155504.79+134335.8$ &  $-401\pm 36 $ & $ -47\pm 23 $ & $ -295\pm 90 $ & \nodata       & \nodata       &  0.71 & $-0.32$ &$-0.58$  &  104 \\ 
$160216.00+111926.9$ &  $-512\pm 49 $ & $-319\pm 39 $ & $ -339\pm 106$ & \nodata       & \nodata       &  0.59 & $-0.24$ &$-0.63$  &  339 \\ 
$163734.52+305403.5$ &  $-634\pm 40 $ & $-813\pm 59 $ & $-1420\pm 47 $ & \nodata       & \nodata       &  0.18 & $-0.77$ &$-1.17$  &    * \\ 
$171304.46+352333.5$ &  $-526\pm 26 $ & \nodata       & \nodata        & \nodata       & \nodata       &  0.25 & $-0.74$ &$-1.42$  &    * \\ 
$205418.80+004915.9$ &  $-412\pm 23 $ & $-423\pm 56 $ & $-1199\pm 181$ & \nodata       & \nodata       &  0.76 & $-0.03$ &$-0.32$  &  309 \\
\enddata

\tablenotetext{\ } {{\it Col 1:}\ Object name.}
\tablenotetext{\ } {{\it Col 2:}\ Velocity offset in [O III]5007 wing component.}
\tablenotetext{\ } {{\it Col 3:}\ Velocity offset in [Ne V]3426 core component.}
\tablenotetext{\ } {{\it Col 4:}\ Velocity offset in [Ne V]3426 wing component.}
\tablenotetext{\ } {{\it Col 5:}\ Velocity offset in [Fe VII]6087 core component.}
\tablenotetext{\ } {{\it Col 6:}\ Velocity offset in [Fe VII]6087 wing component.}
\tablenotetext{\ } {{\it Col 7:}\ Logarithm of [O III]5007/ H$\alpha$ line ratio.}
\tablenotetext{\ } {{\it Col 8:}\ Logarithm of [N II]6583/ H$\beta$ line ratio.}
\tablenotetext{\ } {{\it Col 9:}\ Logarithm of [S II]6716,6731/ H$\beta$ line ratio.}
\tablenotetext{\ } {{\it Col 10:}\ Electron density of NLR derived from [S II] flux ratio. Asterisk (*) represents low density limit ([S II]6716/[S II]6731 $>$ 1.42)}

\end{deluxetable}

\begin{deluxetable}{llllll}
\tablecolumns{6} 
\tablewidth{0pt} 
\tablecaption{Mean Velocity Shift of Coronal Lines}
\tablehead{
&\colhead{} & \multicolumn{2}{c}{Mean $\Delta V$ km s$^{-1}$ (Median)} \vspace{+0.2cm} \\
&\colhead{Coronal line}  &  \colhead{Core component} & \colhead{Wing component}                                          }
\startdata
&[Ne V]3426                & -235$\pm$190 (-205) & -690$\pm$430 (-620)  \\
&[Fe VII]6087              & -340$\pm$190 (-320) & -560$\pm$180 (-540)  \\
&[Fe VII]6087$^*$          & -74$\pm$16          & -296$\pm$28  \\
&[Ne V]3426$+$[Fe VII]6087 & -270$\pm$190 (-220) & -665$\pm$400 (-610)  \\
\enddata

\tablenotetext{*}{Data for stationary AGNs from Gelbord et al. (2009)}

\end{deluxetable} 

\vspace{1.0 cm}

\clearpage
\begin{figure}[h]
\centerline{\includegraphics[scale=0.8]{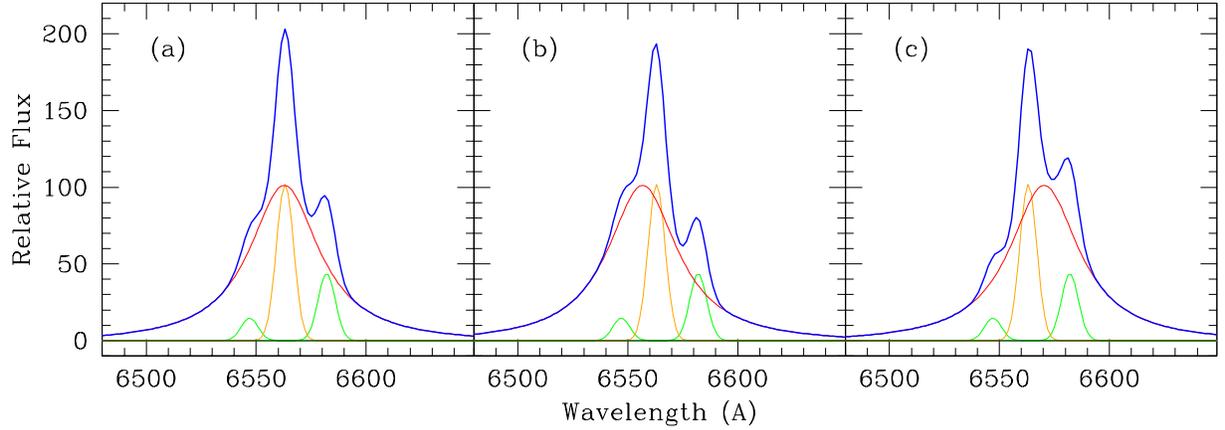}}
\caption{Model spectra of H$\alpha\ +$ [N II] lines (orange \& green: narrow line component of H$\alpha$ \& [N II], respectively (FWHM=400 \kms), red: broad-line component (FWHM=2000 \kms), blue: sum of the narrow$+$broad components). 
Broad-line has no velocity offset (a), broad-line blueshifted (b), and broad-line redshifted (c) with respect to systemic velocity.}
\end{figure}

\begin{figure}[!hb]
\centerline{\includegraphics[scale=0.8]{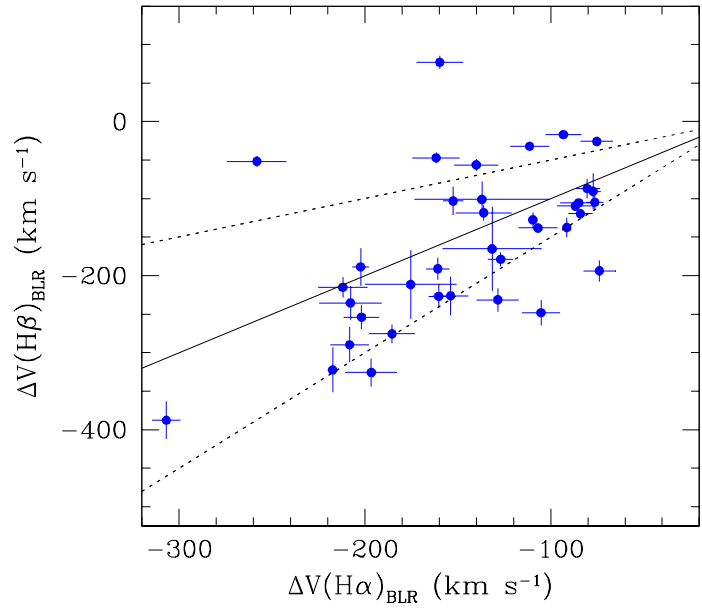}}
\caption{Correlation of velocity offset between H${\alpha}$ and H${\beta}$ broad-lines for the rSMBH candidates.
The solid and dotted lines are one-to-one correspondence and $\pm 50\%$ deviation from the one-to-one correspondence, respectively.
Final 26 rSMBH candidates selected are within the dotted lines.
}
\end{figure}

\begin{figure}[!ht]
\centerline{\includegraphics[scale=0.9]{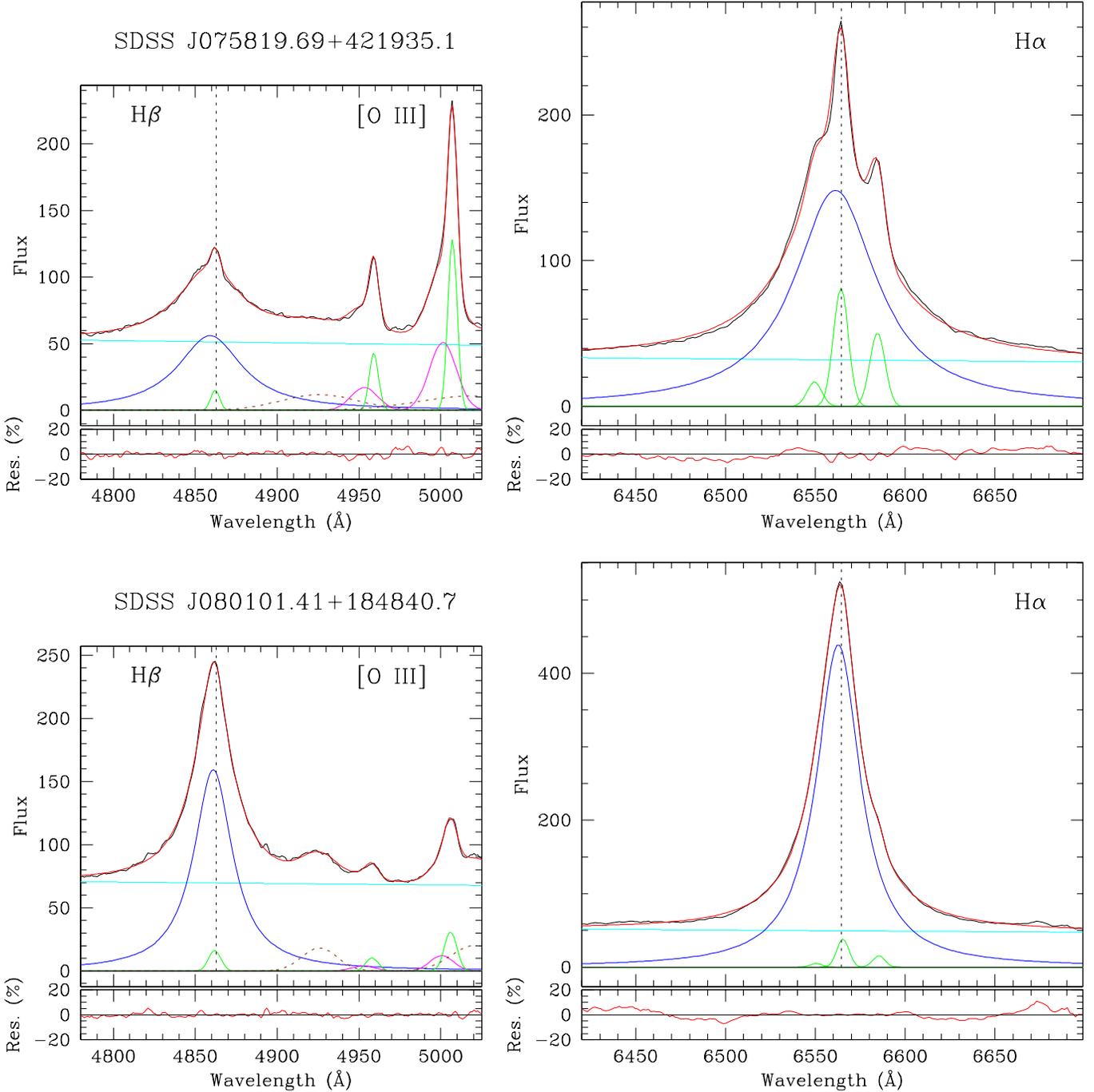}}
\caption{Result of spectral decompositions for H${\alpha}$ and H${\beta}$ ($+$ [O III]) lines,
where black, cyan, blue, green, magenta, and red lines represent data, power-law continuum, broad emission line, narrow emission lines, [O III] lines, and model, respectively.
Brown dotted lines represent [Fe II] lines and vertical dotted line in each plot represents rest frame line center for H${\alpha}$ and H${\beta}$.
Fitting residual (data/model in percentage) is plotted on the bottom of each plot
(The complete figure set (26 sources) is available in the online journal.)
}
\end{figure}

\begin{figure}[!ht]
\centerline{\includegraphics[scale=0.9]{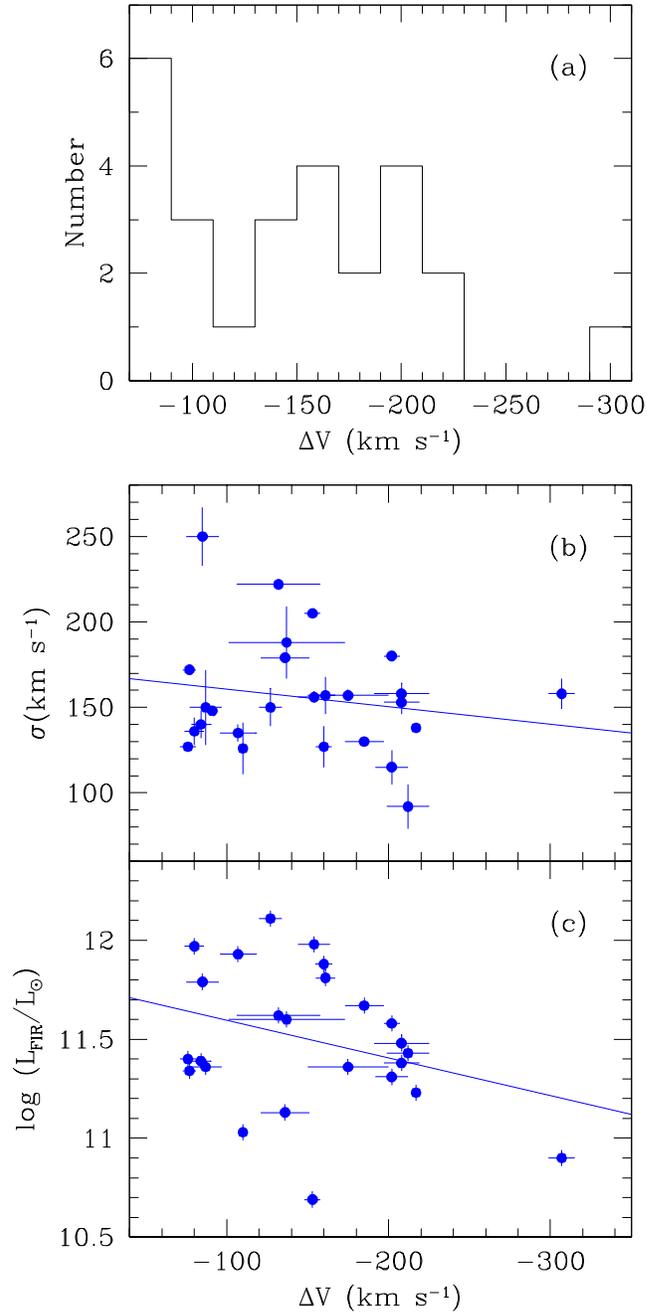}}
\caption{(a) Histogram of projected recoil velocity, (b) recoil velocity vs. velocity dispersion for rSMBH candidates, and (c) recoil velocity vs. infrared luminosity for rSMBH candidates.}
\end{figure}

\begin{figure}[!ht]
\centerline{\includegraphics[scale=0.9]{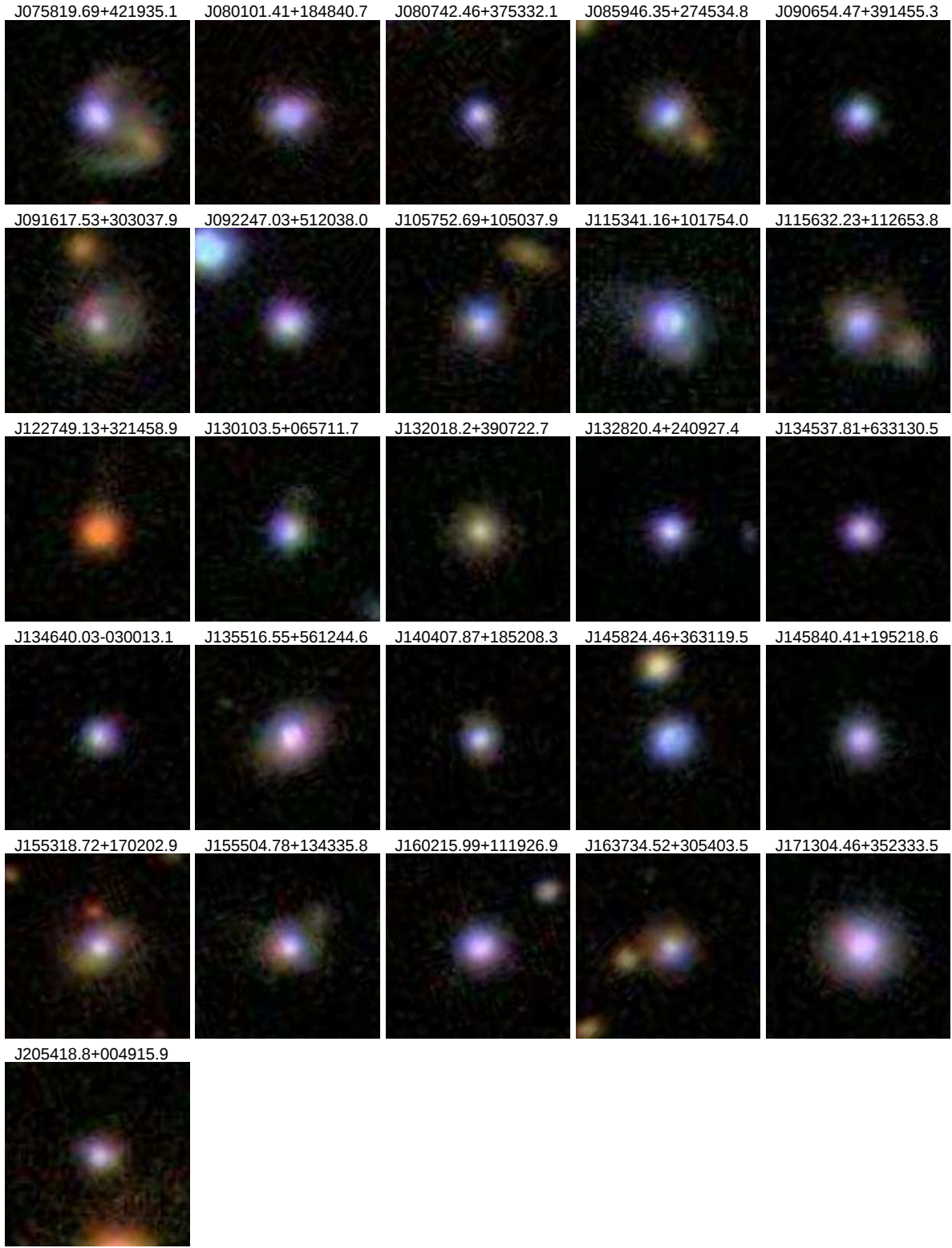}}
\caption{SDSS g-r-i composite image of rSMBH candidates.  Field of view is $20 \arcsec \times 20 \arcsec$.}
\end{figure}

\begin{figure}
\epsscale{1.05}
\plotone{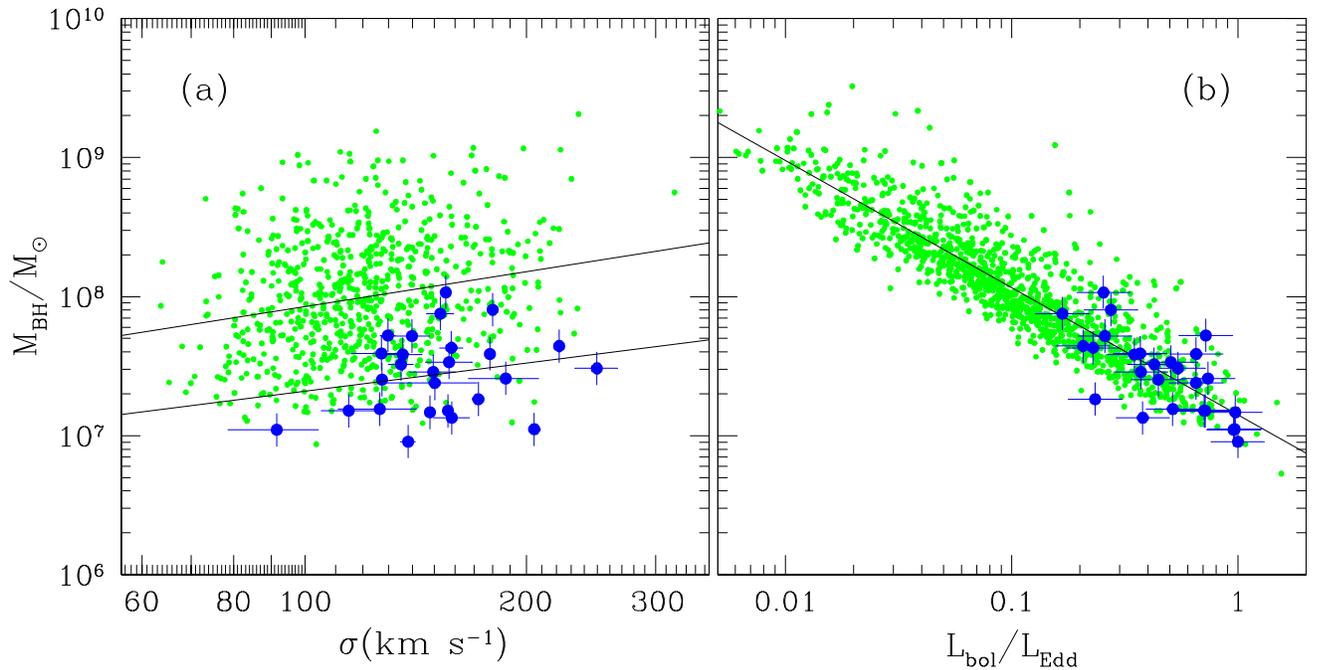}
\caption{(a) M$_{BH}-\sigma_*$ relation and (b) M$_{BH}$ vs. Eddington ratio plots
for rSMBH candidates (blue circles) and SDSS QSOs with z$<$ 0.25 (green dots).
The rSMBH candidates have smaller M$_{BH}$ ($\sim 5 \times$) and larger ($>$0.1) Eddington ratio compared to their stationary counterpart. 
Solid lines represent least-square fit for each data.
}
\end{figure}

\begin{figure}[!ht]
\centerline{\includegraphics[scale=0.9]{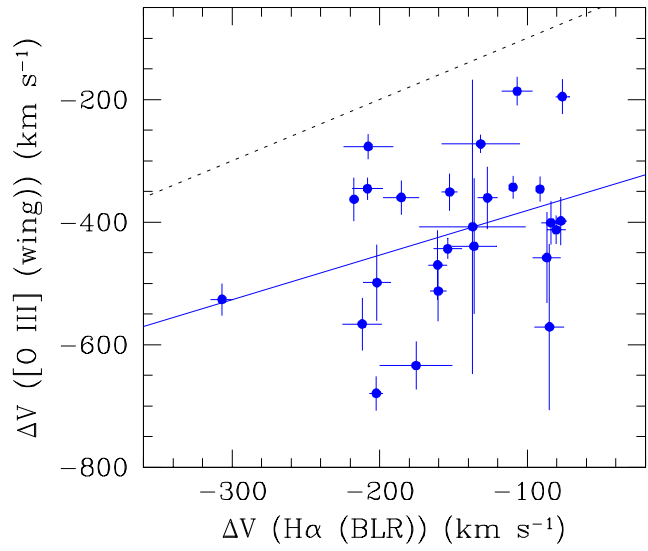}}
\caption{
Velocity offsets in H${\alpha}$ broad-line vs. [O III]$\lambda$5007 wing component,
where dashed and solid lines represent one to one relationship and least-square fit of the data, respectively.
}
\end{figure}

\begin{figure}[!ht]
\centerline{\includegraphics[scale=0.9]{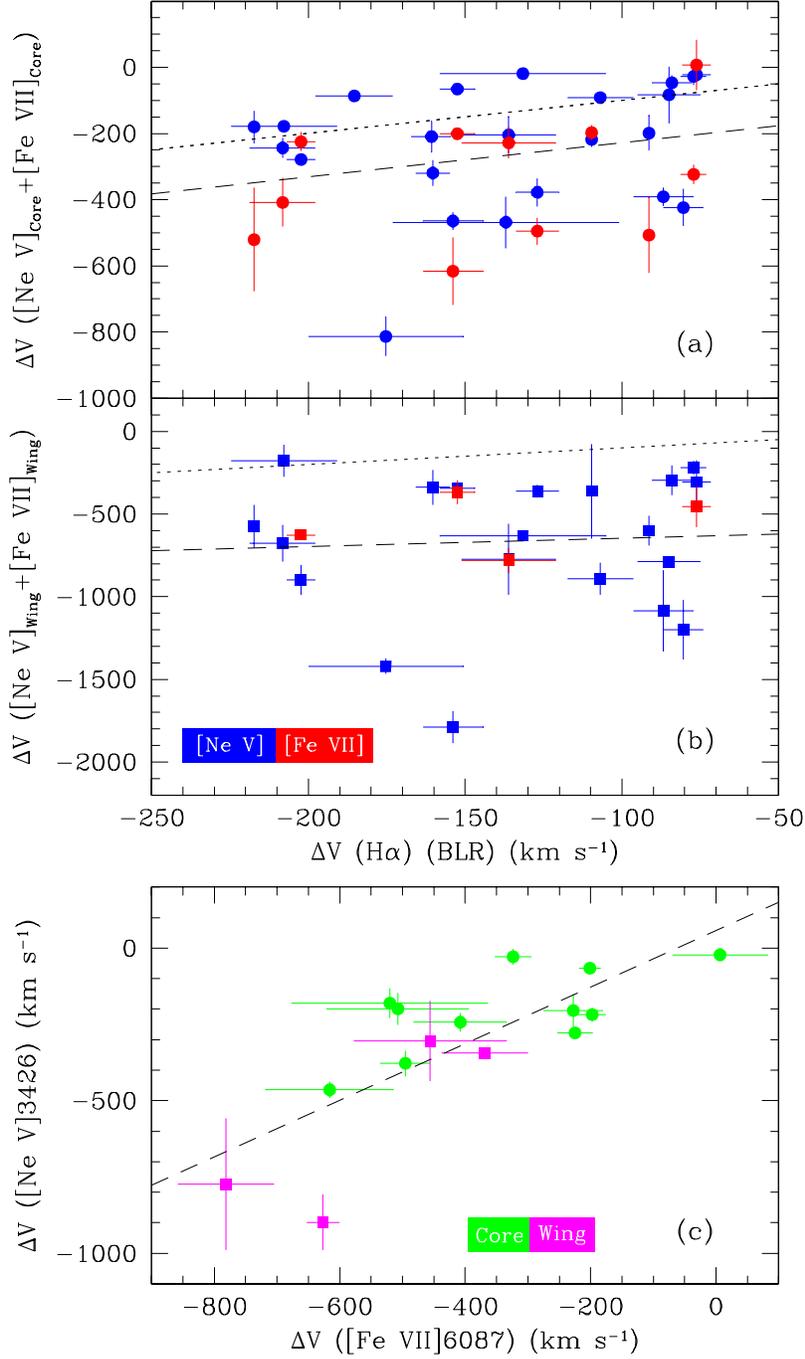}}
\caption{
(a) Velocity offset plot of H${\alpha}$ broad-line vs. [Ne V] (blue circles) \& [Fe VII] (red circles) core components and (b) wing components.
(c) Velocity offset plot of [Ne V] vs. [Fe VII] lines for core (green circles) and wing (magenta circles) components. 
The dotted lines represent one-to-one correspondance and the dashed lines are least-square fit of the data.
}
\end{figure}

\begin{figure}[!ht]
\centerline{\includegraphics[scale=0.9]{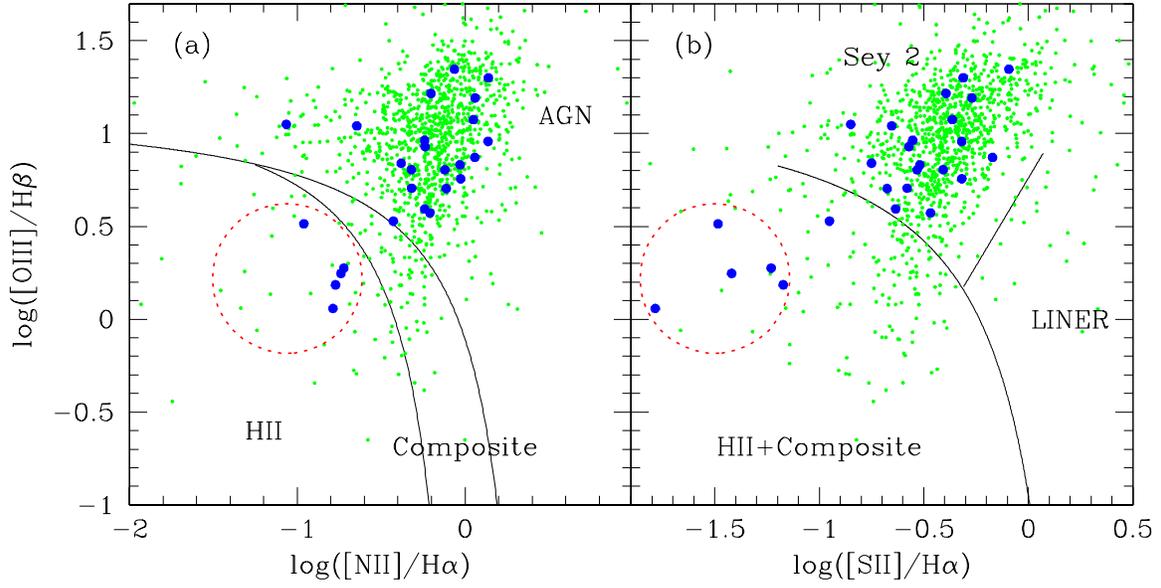}}
\caption{
(a) Narrow line ratio plots of [N II]/H$\alpha$ vs. [O III]/H$\beta$ and (b) [S II]/H$\alpha$ vs. [O III]/H$\beta$,
where blue circles and green dots represent rSMBH candidates and SDSS QSOs, respectively.
Five sources inside red-dotted circle represents H II region rSMBH candidates.
}
\end{figure}

\clearpage

\end{document}